\documentclass[sigconf]{acmart}

\usepackage{multirow}
\usepackage{subfigure}
\usepackage{float}
\usepackage{dsfont}
\usepackage{booktabs}

\usepackage{etoolbox}
\usepackage{graphicx} 
\usepackage{xcolor}
\usepackage{enumitem}
\usepackage{float}
\AtBeginDocument{%
  }

\copyrightyear{2026}
\acmYear{2026}
\setcopyright{cc}
\setcctype{by}
\acmConference[CIKM '26]{Proceedings of the 35th ACM International Conference on Information and Knowledge Management}{November 07--11, 2026}{Rome, Italy}
\acmBooktitle{Proceedings of the 35th ACM International Conference on Information and Knowledge Management (CIKM '26), November 07--11, 2026, Rome, Italy}
\acmDOI{10.1145/3799682.3840983}
\acmISBN{979-8-4007-2539-5/2026/11}

\begin{document}

\title{\textit{SelfDR}: Self-Distillation from Reasoning for LLM-Based Recommendation}

\author{Chumeng Jiang}
\affiliation{%
  \institution{DCST, Tsinghua University}
  \institution{Quan Cheng Laboratory}
  \city{Beijing}
  \country{China}
}
\email{jcm24@mails.tsinghua.edu.cn}

\author{Jiayin Wang}
\affiliation{%
  \institution{DCST, Tsinghua University}
  \city{Beijing}
  \country{China}
}
\email{jiayinwangthu@gmail.com}

\author{Xinjie Lin}
\affiliation{%
  \institution{DCST, Tsinghua University}
  \city{Beijing}
  \country{China}
}
\email{lin-xj22@mails.tsinghua.edu.cn}

\author{Zhiqiang Guo}
\affiliation{%
  \institution{DCST, Tsinghua University}
  \city{Beijing}
  \country{China}
}
\email{georgeguo.gzq.cn@gmail.com}

\author{Hengliang Luo}
\affiliation{%
  \institution{Meituan}
  \city{Beijing}
  \country{China}
}
\email{luohengliang@meituan.com}

\author{Min Zhang}
\authornote{Corresponding author.}
\affiliation{%
  \institution{Quan Cheng Laboratory}
    \institution{DCST, Tsinghua University}
  \city{Beijing}
  \country{China}
}
\email{z-m@tsinghua.edu.cn}
\begin{abstract}
Large Language Models (LLMs) have recently emerged as powerful backbones for recommendation. To better elicit their capabilities, reasoning has been widely incorporated to help LLMs interpret rich textual signals and improve recommendation accuracy. However, explicitly generating intermediate reasoning traces often incurs substantial computational costs, which limits practical deployment in real-world recommender systems.
To address this challenge, we propose \textbf{\textit{SelfDR}}, a \textbf{Self}-\textbf{D}istillation from \textbf{R}easoning framework for LLM-based Recommendation. $SelfDR$ distills an LLM's own reasoning-enhanced predictions to produce recommendations directly, improving recommendation effectiveness while maintaining inference efficiency. All components in the framework are built on the same base LLM, without relying on any external models. Specifically, the teacher recommender is constructed by training a reasoner with downstream performance as the reward, enabling it to generate targeted rationales that are later incorporated into the teacher’s input. A student recommender for direct recommendation, with the same underlying model, then learns from the teacher through self-distillation with a dynamic weighting strategy.
Extensive experiments on three public datasets validate the effectiveness, rationality, and efficiency of $SelfDR$. Codes are available at https://github.com/JiangDeccc/SelfDistillation.

\end{abstract}




\maketitle


\section{Introduction}

In recent years, LLM-based recommendation has attracted increasing attention and has demonstrated promising results across a variety of tasks~\cite{llmrec, benefit, survey, llmreceva}. A wide range of mechanisms have been created to adapt the capabilities of LLMs to recommendation settings, including in-context learning~\cite{wang2023zeroshot}, supervised fine-tuning~\cite{bao2023bistep, openp5} and so on. These approaches shed light on the potential of aligning the recommendation representation space with the semantic space, thereby harnessing the capabilities of LLMs to enhance recommendation performance.

\begin{figure}[ht]
\setlength{\abovecaptionskip}{0.2cm}
\setlength{\belowcaptionskip}{-0.2cm}
  \centering
  \includegraphics[width=\linewidth]{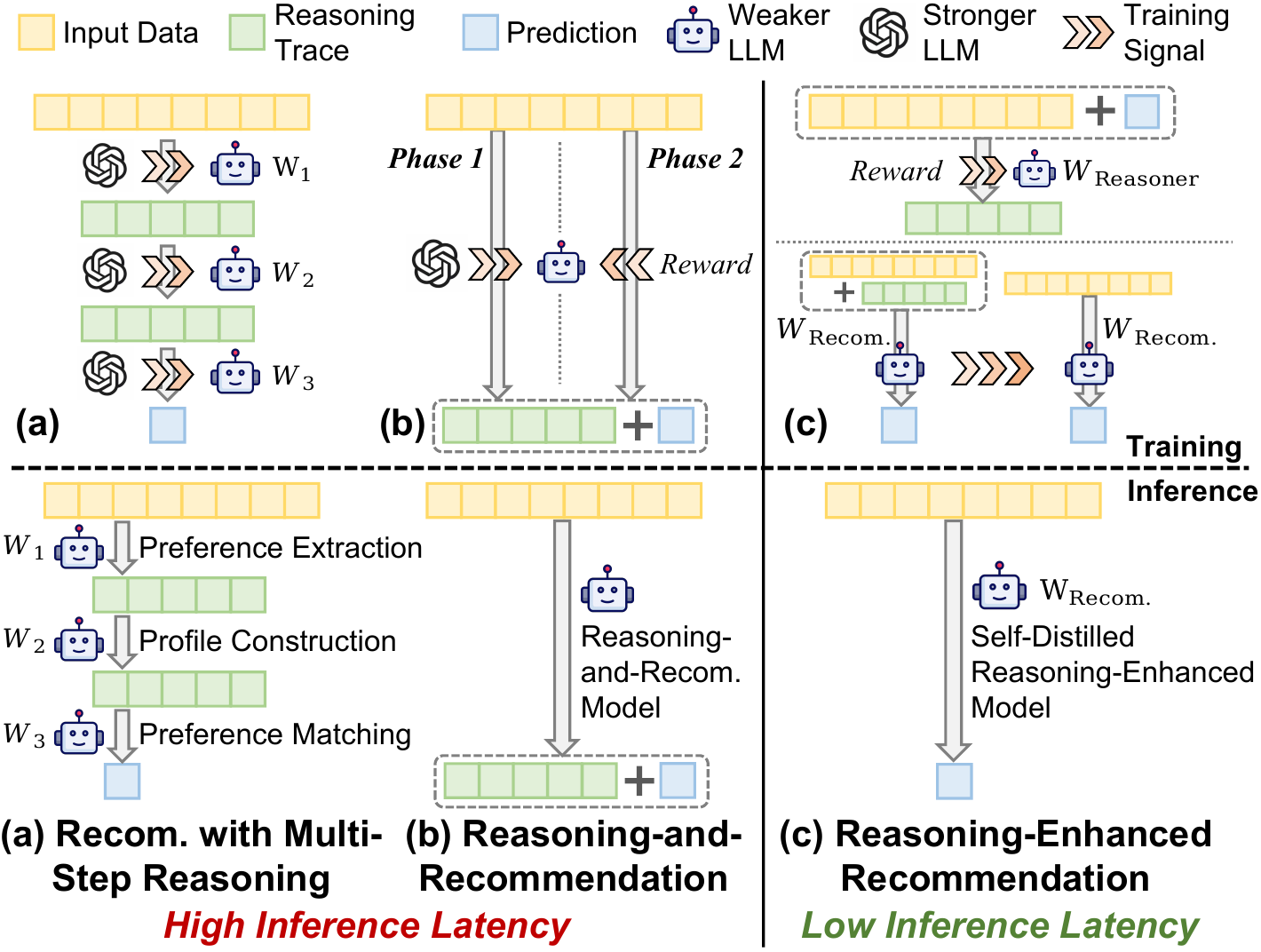}
  \caption{Two previous typical reasoning paradigms: (a) multi-step reasoning and (b) reasoning-and-recommendation. 
  Our proposed paradigm (c) self-distills from reasoning to enhance effectiveness while ensure inference efficiency.
  }
  \label{fig:intro}
\end{figure}

Furthermore, to help LLMs better understand the vast textual information in recommendation data, researchers have begun to explore the use of reasoning to enhance recommendation effectiveness, motivated by the success of incorporating explicit reasoning in general LLMs~\cite{dotamath, imani-etal-2023-mathprompter}. One line of work follows a multi-step reasoning paradigm (Figure~\ref{fig:intro}(a)), where the model sequentially reasons over user and item information and ultimately leverages the resulting profiles to support recommendation~\cite{expert, fang2025reason4reclargelanguagemodels}. Another direction adopts a reasoning-and-recommendation paradigm (Figure~\ref{fig:intro}(b)), which performs both within a single inference step by encouraging the generation of long reasoning traces or reasoning tokens before the final output, thereby refining predictions~\cite{RecSAVER, bismay-etal-2025-reasoningrec, slowthinking}. These methods have achieved strong performance, highlighting the potential of reasoning-augmented recommendation.

However, such paradigms inevitably impose substantial inference overhead, which is often unacceptable in real-world recommender systems where responsiveness is crucial. To address this challenge, we argue for a paradigm that continues to focus on the \emph{direct recommendation} (Figure~\ref{fig:intro}(c))—given the necessary inputs, such as user history, the model directly outputs the recommendation results. This design minimizes inference costs and improves practical usability.
Nevertheless, though we hope to maintain the brevity of the outputs, reasoning itself can be beneficial to improve the recommendation quality, as it helps LLMs to better interpret and organize the rich information in recommendation data. In this case, it raises the key research question of how to exploit the reasoning capability of LLMs without sacrificing inference efficiency.


To tackle this, we propose a \textbf{Self}-\textbf{D}istillation from \textbf{R}easoning framework for LLM-based Recommendation (\textbf{\textit{SelfDR}}), which distills the model’s own deliberated predictions to enable reasoning-enhanced recommendation without requiring explicit reasoning. To achieve this, \textit{first}, we construct a teacher model equipped with an additional deliberation step. Specifically, we train a reasoner to generate high-quality rationales based on the user history and the next interacted item in the training set, which are then used as auxiliary input for a reasoning-enhanced teacher that produces more effective and deliberative responses.
\textit{Next}, we let a student recommender LLM, which performs direct recommendation, learn from the teacher and internalize the outputs enhanced by the deliberation process. To ensure effective learning, a dynamic weighting mechanism is adopted to adapt the learning signals. Throughout the entire pipeline, the reasoner and the recommenders share the same underlying architecture.
Our extensive experiments evaluate the effectiveness, rationality, and efficiency of $SelfDR$ in various scenarios. The $SelfDR$ reasoner achieves superior downstream performance compared to external large LLMs and after distillation, our method consistently delivers the best recommendation results with the lowest inference cost across all three datasets.

Our main contributions are summarized as follows:
\begin{itemize}


    \item We propose $SelfDR$, a self-distillation framework that enhances recommendation quality through self-improvement without relying on stronger external LLMs.

    \item Unlike previous self-distillation settings, $SelfDR$ distills from the reasoning process, thereby reducing the need for explicit intermediate reasoning outputs and improving both effectiveness and efficiency.

    \item Extensive experiments on three public datasets demonstrate the effectiveness, rationality, and efficiency of $SelfDR$, showcasing the potential of reasoning-guided distillation to facilitate the self-evolution of LLMs.

\end{itemize}
\section{Related Work}
\subsection{LLM-Based Recommendation}

Recent advances in large language models (LLMs) have driven their adoption in recommender systems. Existing approaches align LLMs with recommendation through (i) zero-shot prompting~\cite{liu2023chatgpt, liu2023llmrec, wang2023zeroshot}, (ii) fine-tuning on recommendation data~\cite{bao2023bistep, openp5}, and (iii) hybrid architectures that incorporate collaborative-filtering signals~\cite{lcrec, transrec, collm}. By leveraging LLMs’ language understanding and broad knowledge, these methods improve recommendation performance and help address challenges such as cold start~\cite{TALLRec, ALLMRec}.

However, preference signals embedded in item metadata and user-generated text, such as reviews, are often subtle and difficult to interpret reliably. To help LLMs better exploit such information, inspired by the success of chain-of-thought reasoning in general tasks~\cite{imani-etal-2023-mathprompter, dotamath, xu-etal-2024-faithful}, recent studies introduce explicit reasoning into LLM-based recommendation. One line of work adopts multi-step pipelines that iteratively reason over user and item text. For example, EXP3RT~\cite{expert} decomposes recommendation into preference extraction, profile construction, and reasoning-enhanced rating prediction, with a dedicated model trained for each stage. Often trained with detailed feedback, such methods improve both accuracy and explainability in rating prediction~\cite{fang2025reason4reclargelanguagemodels} and ranking~\cite{cot4rec}.
Another line of work employs single-step test-time reasoning, generating intermediate reasoning traces before predictions to identify and revise suboptimal outputs. Some methods~\cite{RecSAVER, bismay-etal-2025-reasoningrec} elicit chain-of-thought explanations for personalization, while others~\cite{slowthinking, recr1} use reinforcement learning to acquire effective reasoning patterns from historical interactions.

Despite their effectiveness, reasoning-based approaches incur substantial computational overhead. Multi-step pipelines and test-time reasoning traces increase latency and resource consumption, limiting their practicality in large-scale, time-sensitive recommender systems. In contrast, our work targets direct recommendation, producing results in a single pass while transferring the benefits of LLM reasoning to an efficient inference process. This design balances recommendation effectiveness with practical efficiency.

\begin{figure*}[h]
\setlength{\abovecaptionskip}{0.1cm}
\setlength{\belowcaptionskip}{0cm}
  \centering
  \includegraphics[width=0.95\linewidth]{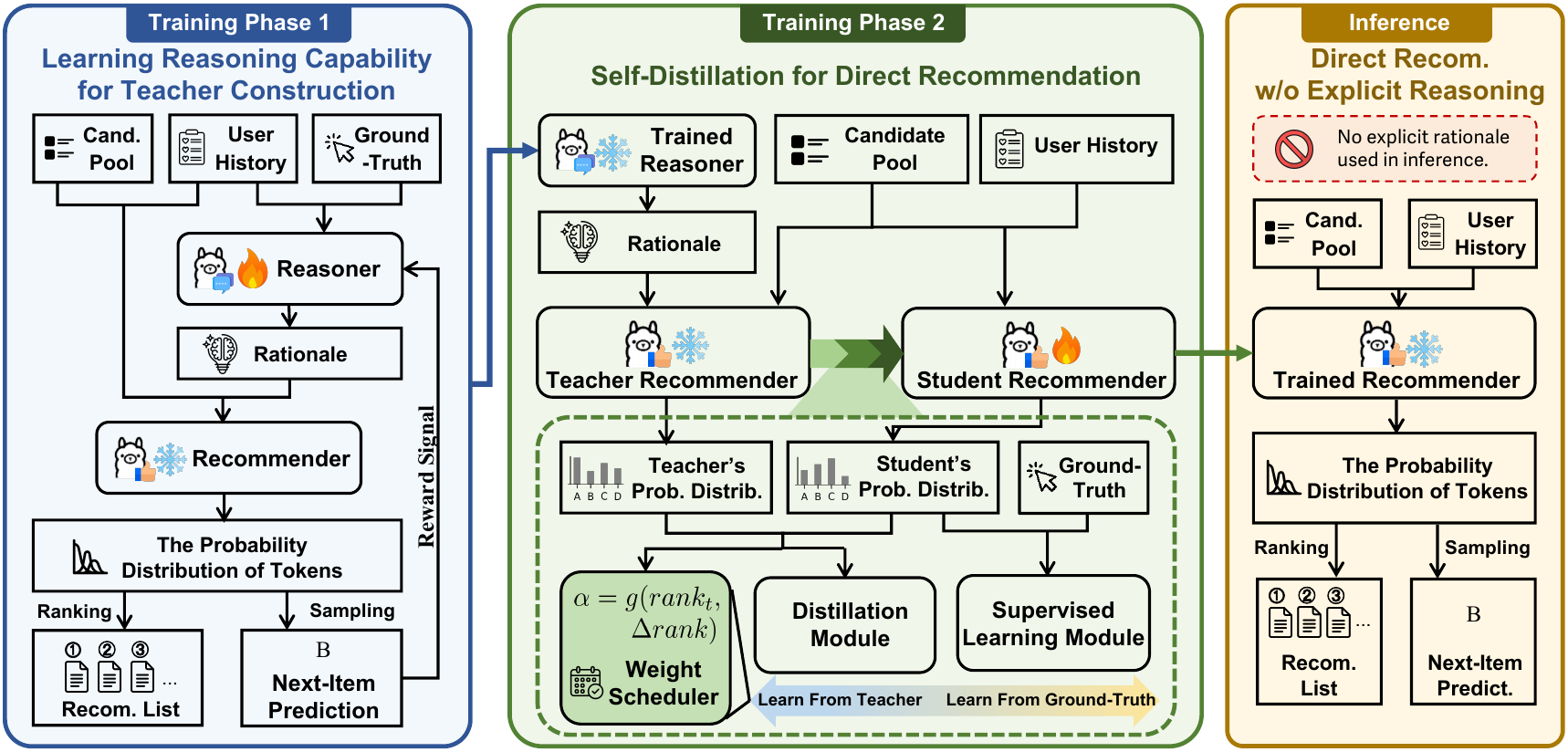}
  \caption{The $SelfDR$ framework. A reasoner is trained with recommendation performance as the reward to generate high-quality rationales for a rationale-augmented teacher. A direct-recommendation student then learns from the teacher’s deliberative outputs for reasoning-enhanced ranking without explicit rationales. All LLMs share the same base model.
  }
  \label{fig:overview}
\end{figure*}

\subsection{Knowledge Distillation}

Knowledge distillation (KD) transfers knowledge from a high-capacity teacher model to a lightweight student~\cite{kd}, and has been widely adopted in computer vision (CV)~\cite{habib2024comprehensivereviewknowledgedistillation} and natural language processing (NLP)~\cite{xu2024surveyknowledgedistillationlarge} for model compression and acceleration. The key idea is to have the student mimic the teacher’s behavior, whether by matching its final predictions (\textit{response-based})~\cite{lkd}, internal feature representations (\textit{feature-based})~\cite{romero2015fitnetshintsdeepnets, featurekd}, or relational structures among samples or layers (\textit{relation-based})~\cite{rkd}.

In recommender systems, KD is commonly used to reduce representation dimensionality and inference latency, typically by compressing a deep model into a shallower one~\cite{hetcomp, featurekdrs, sang2024featureinteractionfusionselfdistillation}. With the rise of LLMs, KD has further been used to transfer their semantic reasoning into traditional recommenders, mitigating high inference costs~\cite{wu2025bidirectionalknowledgedistillationenhancing}: DLLM2Rec~\cite{dllmrec} combines importance-weighted ranking with collaborative embedding distillation to filter unreliable signals and align semantic spaces. Another paradigm instead distills a large LLM into a smaller, recommendation-oriented LLM—SLMRec~\cite{xu2025slmrec} prunes redundant layers while retaining performance, and ALKDRec~\cite{alkdrec} actively queries the teacher on informative samples to fine-tune the student for session recommendation. Overall, distilling LLM knowledge, whether into conventional recommenders or smaller LLMs, helps balance accuracy and efficiency.
These approaches, however, face two key limitations. First, they depend on powerful external LLMs: if the teacher is inaccessible or poorly aligned with the target domain, distillation may underperform, and better teachers do not always yield better students, making teacher selection non-trivial~\cite{dong2024toward}. Second, keeping large LLMs in the training loop is computationally expensive.

The above limitations motivate the exploration of self-distillation techniques. Self-distillation leverages the model's own knowledge as the teaching signal, where a model's deeper-layer features~\cite{yang2023knowledge} or an ensemble of its snapshots~\cite{yang2019snapshot} can serve as the teacher for the same model at a later stage. 
It has been successfully applied in CV~\cite{selfkdcv, bahmani2025lyragenerative3dscene} and NLP~\cite{hahn-choi-2019-self} to improve model generalization and performance without extra model overhead. 
In recommendation, prior studies have also explored self-distillation for conventional recommender models, such as enriching user preference representations by retrieving informative interactions from similar users as auxiliary self-supervision signals~\cite{llmesr}.
More recently, SOFT explores LLM self-distillation by treating the outputs of a fine-tuned model as auxiliary easy-to-learn data to improve optimization on real recommendation data~\cite{soft}.
Despite these encouraging efforts, self-distillation, especially from reasoning process, remains largely unexplored in LLM-based recommendation.


\section{Methodology}

\subsection{Overview}
While introducing explicit reasoning steps can improve recommendation quality, it also incurs substantial inference costs. To balance these factors, we propose \textbf{$SelfDR$}, a self-distillation framework that uses the model's own deliberation-enhanced supervision to train a direct-recommendation (Sec.~\ref{derrec}) model. Specifically, we first construct a reasoning-enhanced teacher model (Sec.~\ref{teacher}) following the deliberative recommendation paradigm (Sec.~\ref{delrec}): in the reason-only generation stage, we train a reasoner with the downstream performance as reward to generate targeted explanations, which are then incorporated into the recommendation process to form the teacher recommender. Once the teacher is established, a direct-recommendation student model sharing the same underlying architecture as the teacher is trained to learn from the teacher’s deliberative outputs with dynamic loss (Sec.~\ref{curriculum}), guiding the student to acquire reasoning-enhanced ranking behavior without relying on explicit rationales. The entire learning and evolution process can be accomplished by a same backbone LLM without relying on external LLMs.

\subsection{Task Formulation}
\subsubsection{Direct Recommendation}
\label{derrec}
We define the direct recommendation task as producing the final recommendation output directly from the necessary input information (e.g. user history and candidate set) without any intermediate steps. This is the task that we ultimately expect our model to perform, as it better ensures inference efficiency, which is often critical in recommender systems.

Formally, let $\mathcal{U}$ denote a set of users and $\mathcal{I}$ a set of items.  
Each item $i \in \mathcal{I}$ is associated with metadata, such as title and category, denoted by $\mathcal{M}_i$.  
The user history $H_u = (h_{u,1}, h_{u,2}, \ldots, h_{u,t})$ constitutes a chronologically ordered sequence of past records.  
Each record is defined as $h_{u,i} = (\mathcal{M}_i, d_{u,i})$, where $d_{u,i}$ denotes the interaction details of user $u$ with item $i$, e.g., rating. The primary objective of the \emph{direct recommendation} is to forecast the next item the user is most likely to interact with from the candidate set.  
This can be formalized as follows:
\begin{equation}
\mathcal{T}_{u,i} = f(H_u, \mathcal{C}_{u,i}),
\end{equation}
where $\mathcal{C}_{u,i} \subseteq \mathcal{I}$ denotes the candidate set for user $u$ with respect to the ground-truth item $i$. 
It can be constructed under either ranking or reranking settings: the former typically combines $i$ with randomly sampled negatives, while the latter uses the top-$K$ items retrieved by a base ranker based on user $u$'s interaction history prior to item $i$, where ground-truth item $i$ is not necessarily included.

When employing an LLM for the task, we follow the prompting strategy adopted in prior studies\cite{COTRec, llamarec}, where the LLM is instructed to predict the unique character identifier of the item which the user is most likely to engage with. In this paradigm, we obtain both a single generated next-item prediction $s_{u,i}$ and a recommendation list $\mathcal{R}_{u,i}$ formed by ranking candidate items according to the LLM-generated log-probabilities of their index tokens. The task can now be formulated as:
\begin{equation}
s_{u,i}, \mathcal{R}_{u,i} = LLM(H_u, \mathcal{C}_{u,i}).
\end{equation}

\subsubsection{Deliberative Recommendation}
\label{delrec}
In contrast to direct recommendation, the \emph{deliberative recommendation} incorporates intermediate stages of reasoning, where the model engages in understanding and reflecting on user history or item information before making a prediction. This paradigm extends the reasoning chain for improved effectiveness while inevitably increasing inference time. 

In this paper, our teacher recommender adopts this task paradigm on the training set to enhance performance through reasoning. To obtain a more effective teacher model, we decompose the task into two stages: \emph{reason-only generation} and \emph{reason-enhanced recommendation}.
In the reason-only generation stage, the Reasoner LLM takes the user history and a ground-truth item as input and is prompted to infer why the user is likely to engage with the item.
Specifically, it generates an explicit reasoning trace given both the question and the answer, producing more substantive and informative reasoning content that subsequently enhances the teacher’s recommendation quality. Formally, this step can be expressed as:
\begin{equation}
r_{u,i} = LLM_{Reasoner}(H_u, \mathcal{M}_{i}),
\label{eq:reasoner}
\end{equation}
where $r_{u,i}$ represents the rationale for the user’s preference for the next item $i$. Here we also adopt the prompting strategy that has been used in previous work~\cite{RecSAVER}.  

Subsequently, the reason-enhanced recommendation stage is performed. In this stage, the input incorporates not only the user history and the candidate set, but also the previously generated reasoning content, which captures both the model’s understanding of the user’s past interactions and the characteristics of items the user is likely to engage with in the future, thereby assisting the LLM in producing more accurate recommendations:
\begin{equation}
s_{u,i}, \mathcal{R}_{u,i} = LLM(H_u, \mathcal{C}_{u,i}, r_{u,i}).
\label{eq:teacher}
\end{equation}

\subsection{Learning Reasoning Capability for Teacher Construction}
\label{teacher}

We construct the teacher recommender LLM under the deliberative recommendation paradigm, which explicitly leverages the reasoning capability of LLMs. In this setting, the first step is to obtain a Reasoner capable of generating explanations that are sufficiently targeted and useful for the recommendation task. A straightforward approach might be using the user’s reviews as reasons. However, not all datasets contain user reviews, and when reviews are available, they often include substantial irrelevant information, leading to inferior performance. 

In this case, we employ Group Relative Policy Optimization (GRPO)~\cite{GRPO}, using the accuracy of the downstream task as the reward signal to train a Reasoner based on the same backbone LLM to produce more helpful explanations.
GRPO is a reinforcement learning algorithm that offers more stable convergence and lower memory overhead, making it suitable for large-scale LLM training. 
Further, using the downstream recommendation performance as the reward signal avoids potential mismatch between separately defined reasoning evaluation metrics and the ultimate recommendation objective, thus ensuring that the generated reasoning can be aligned with the end task. 
Concretely, given the user history $H_u$ and ground-truth item information $\mathcal{M}_i$, the Reasoner generates a reason $r_{u,i}$ (Eq.~\ref{eq:reasoner}), which is then combined with the user history $H_u$ and candidate set $C_{u,i}$ and fed into another frozen recommender LLM to perform reason-enhanced recommendation (Eq.~\ref{eq:teacher}). The correctness of the generated next-item prediction $s_{u,i}$ serves as the reward signal for guiding the Reasoner’s explanation generation:
\begin{equation}
Reward(r_{u,i}) = \mathrm{Acc}_{rec}(s_{u,i}),
\end{equation}
where $\mathrm{Acc}_{rec}$ can be instantiated as any metric that evaluates recommendation accuracy. 

To make the reward better aligned with the intrinsic quality of the generated rationale, we explicitly prompt the Reasoner not to directly restate the input metadata, thereby mitigating direct information leakage. Moreover, before incorporating $r_{u,i}$ into the recommender input, we apply a fine-grained masking step: if the generated rationale contains three or more consecutive words that exactly overlap with the item title, these overlapping words are masked before reward computation.

With the trained Reasoner, we instruction-tune a recommender LLM and concatenate the reasoning content generated by the Reasoner with the recommendation input, forming the teacher LLM used for subsequent distillation.

\subsection{Self-Distillation for Direct Recommendation}
\label{curriculum}

The ultimate goal of $SelfDR$ is to leverage the reasoning capacity to perform the direct recommendation more effectively. To this end, the next step is to enable a student LLM, which is designed for direct recommendation, to learn from the deliberative output of the teacher LLM, which incorporates reasoning as part of its input. Before distillation, both the teacher and student recommender LLMs are first instruction-tuned to familiarize them with the recommendation task and enforce instruction adherence; at this stage, the student and teacher share identical model parameters.

During the self-distillation process, the key distinction between the two lies in the input: while the teacher recommender LLM additionally receives the Reasoner-generated rationale, the student recommender LLM does not:
\begin{equation}
\begin{split}
   & Teacher: s^t_{u,i}, \mathcal{R}^t_{u,i} = f(H_u, \mathcal{C}_{u,i}, r_{u,i}), \\
   & Student: s^s_{u,i}, \mathcal{R}^s_{u,i} = f(H_u, \mathcal{C}_{u,i}).
\end{split}
\end{equation}
where $f(\cdot)$ indicates the instruction-tuned recommender LLM.
The purpose of this design is to allow the student to learn from the teacher’s deliberated outputs and gradually develop stronger recommendation capabilities. It is also worth noting that concise outputs keep the training cost modest compared to methods relying on long reasoning-based generations.

In this case, the most valuable supervision from the teacher lies in its logits, since simply learning from its generated outputs may often be equivalent to learning the labels, causing the process to degenerate into standard fine-tuning. Therefore, we align the student’s predictive distribution with the teacher’s using the logits before the softmax layer, thereby providing a smoother and more informative training signal. Formally, the optimization objective can be expressed as:
\begin{equation}
\min_{\Theta_s} \big[(1-\alpha)*\mathcal{L}_{ce}(\Theta_s) + \alpha* \mathcal{D}_{kd}(\Theta_t, \Theta_s) \big],
\end{equation}
where $\mathcal{L}_{ce}$ denotes the cross-entropy loss with respect to the ground-truth, and $\mathcal{D}_{kd}(\Theta_t, \Theta_s)$ represents the distillation loss measuring the divergence between the logits of the teacher and the student. Our method follows an offline distillation scheme, keeping the teacher recommender LLM fixed during the distillation process.
For the distillation objective, we employ the reverse KL divergence~\cite{gu2024minillm} as the loss function, which tends to encourage the student distribution to concentrate on the high-probability regions of the teacher distribution. Accordingly, the distillation loss can be formulated as:
\begin{equation}
\begin{split}
    \mathcal{D}_{kd}(\Theta&_t, \Theta_s) = \mathcal{L}_{\text{RKL}}(\Theta_s \mid \Theta_t) \\
=\sum_{j \in \mathcal{C}_{u,i}} & p_{\Theta_t}(\mathcal{CI}_j \mid H_u, \mathcal{C}_{u,i}, r_{u,i}) \, \log \frac{p_{\Theta_t}(\mathcal{CI}_j \mid H_u, \mathcal{C}_{u,i}, r_{u,i})}{p_{\Theta_s}(\mathcal{CI}_j \mid H_u, \mathcal{C}_{u,i})} ,
\end{split}
\end{equation}
where $p_{\Theta_t}$ and $p_{\Theta_s}$ denote the teacher and student’s output distributions over the character identifiers of candidates $\mathcal{CI}_{j}$, respectively, given the user history $H_u$ with or without rationale $r_{u,i}$.

To make the self-distillation process more effective, we introduce a dynamic weighting scheme that adaptively adjusts the learning weight $\alpha$ of the distillation loss, because the teacher, which incorporates Reasoner-generated rationale, does not always produce reliable or fully learnable outputs despite providing valuable guidance.
The proposed dynamic weighting mechanism therefore balances supervision from the teacher and the ground-truth labels, assigning greater emphasis to the teacher when its predictions are more informative, and adapting automatically as training progresses.

There are two primary reasons why the teacher’s outputs may be less instructive.
First, the quality of the teacher’s own predictions may be suboptimal. While we expect the positive item to be ranked within the top-$k$ positions, the teacher may fail to do so. We therefore design an adjustment function based on the distance between the teacher-ranked position of the positive item and a threshold rank $k$ to modulate the learning weight accordingly.
Second, the teacher may underperform the student on certain samples. In such cases, the learning weight from the teacher should be substantially reduced compared to instances where the teacher performs better. To this end, we introduce a decay function based on the rank difference between the teacher and the student, together with an additional penalty parameter to handle underperforming-teacher cases.
The overall function is defined as:

\begin{equation}
    \begin{split}
&\alpha(rank^t_i, rank^s_i)  \\ &=\alpha_{base}f(rank^t_i-k)\sigma(\dfrac{rank^s_i-rank^t_i}{\tau})\beta^{\ \vmathbb{1} \left(rank_i^t > rank_i^s\right)}, \\[0.5em]
&where \ f(x) = 1+ tanh(x/\gamma).
    \label{KS} 
\end{split}
\end{equation}
Here, $rank^t_i$ and $rank^s_i$ denote the ranks predicted by the teacher and student respectively, $\sigma(\cdot)$ is the sigmoid function, $\tau$ and $\gamma$ are temperature coefficients.
To illustrate, when the teacher produces a worse ranking than the student, we regard this case as not beneficial for distillation and apply a penalty term controlled by $\beta$ to substantially reduce $\alpha$. Meanwhile, the weighting function also considers the teacher rank relative to the threshold and the magnitude of the rank difference, for which we adopt soft gating functions to smoothly adjust the reliance on teacher guidance rather than impose hard selection.
This design allows the optimization objective to dynamically emphasize teacher guidance, improving both robustness and effectiveness of the self-distillation process.

\section{Experiments}


\begin{description}[leftmargin=0pt,labelsep=0.0em]
    \item \textit{RQ1:} How does $SelfDR$ perform compared with existing baseline methods? 
    \item \textit{RQ2:} How does the performance of the $SelfDR$ Reasoner compare with that of an external large LLM Reasoner?
    \item \textit{RQ3:} How does self-distillation contribute to recommendation performance?
    \item \textit{RQ4:} What are the training cost and inference efficiency of $SelfDR$ relative to other methods?
\end{description}

\subsection{Experimental Settings}
\subsubsection{Datasets.}
We conduct experiments on three common real-world datasets, \textit{Amazon Clothing Shoes and Jewelry \textbf{(Clothing)}}, \textit{Amazon Home and Kitchen \textbf{(Home)}} and \textit{Movielens-1M\textbf{ (ML1M)}}. The Amazon datasets record user–item interactions across multiple sub-categories on the Amazon platform, accompanied by detailed product metadata and user reviews. ML1M contains over one million movie ratings along with rich user information and movie metadata. 
For data preprocessing, we adopt the widely used 5-core filtering and leave-one-out splitting strategy.
The detailed statistics of these datasets are presented in Table~\ref{Dataset}.

\begin{table}[h]
\setlength{\abovecaptionskip}{0.0cm}
\setlength{\belowcaptionskip}{-0.2cm}
\centering
\caption{Statistics of the experimental datasets.}
\begin{tabular}{lrrrr}
\toprule
      & \#User & \#Item & \#Inter & Density (\%)\\
\midrule
Clothing & 39,387 & 23,033  & 278,677  & 0.0307    \\
Home & 66,519 & 28,237 & 551,682 & 0.0294 \\
ML1M & 6,040 & 3,416 & 999,611 & 4.8448 \\
\bottomrule
\end{tabular}
\label{Dataset}
\end{table}


\begin{table*}[!ht]
\setlength{\abovecaptionskip}{0cm}
\caption{The Recommendation List Quality of $SelfDR$ compared with traditional models and LLM-based recommenders on three datasets. All LLM-based methods utilizing LLaMA-3.1-8B-Instruct as backbone. Bold and underlined indicate the best and the second-best performance, respectively. *($p$-value < 0.05 compared with best baseline).}
\label{tab:main}
    \centering
    \resizebox{\linewidth}{!}{

    \begin{tabular}{cc|c|ccc|cc|cc|cc|cc|c}
    \toprule
        ~ & ~ & ~ & \multicolumn{5}{c|}
        {\textbf{Traditional Recommenders}} & \multicolumn{6}{c}{\textbf{LLM-Based Recommenders}} \\ 
        ~ & ~ & ~ & \multicolumn{3}{c|}
        {\textbf{ID}} & \multicolumn{2}{c|}{\textbf{Content}} & \multicolumn{2}{c|}{\textbf{No Reason.}} & \multicolumn{2}{c|}{\textbf{Multi-Step}} & \multicolumn{2}{c|}{\textbf{Reason.-and-Recom.}} & \textbf{Ours} \\ 
        ~ & ~ & Base Rank & PRM & SetRank & MIR & BGE & Jina & ZSRanker & SOFT & EXP3RT & COT4Rec & RecSAVER & RecR1 & $SelfDR$ \\ \midrule

        \multirow{5}{*}{\rotatebox{90}{Clothing}} & $H_l @1$ &         0.0047  & 0.0063  & 0.0064  & 0.0072  & 0.0079  & 0.0069  & 0.0047  & 0.0113  & 0.0058  & 0.0112  & 0.0095  & \underline{0.0114}  & \textbf{0.0132*}  \\ 
        ~ & $H_l @3$ & 0.0173  & 0.0182  & 0.0182  & 0.0198  & 0.0226  & 0.0213  & 0.0179  & \underline{0.0240}  & 0.0153  & 0.0238  & 0.0180  & 0.0204  & \textbf{0.0249*}  \\
        ~ & $N_l @3$ &  0.0120  & 0.0130  & 0.0131  & 0.0144  & 0.0164  & 0.0151  & 0.0122  & \underline{0.0187}  & 0.0113  & 0.0186  & 0.0144  & 0.0166  & \textbf{0.0200*}  \\ 
        ~ & $H_l @5$ & 0.0251  & 0.0275  & 0.0273  & 0.0291  & 0.0313  & 0.0297  & 0.0263  & 0.0314  & 0.0209  & \underline{0.0315}  & 0.0239  & 0.0263  & \textbf{0.0318*}  \\ 
        ~ & $N_l @5$ & 0.0152  & 0.0169  & 0.0168  & 0.0182  & 0.0200  & 0.0186  & 0.0157  & \underline{0.0217}  & 0.0136  & \underline{0.0217}  & 0.0168  & 0.0191  & \textbf{0.0228*}  \\  \midrule

        \multirow{5}{*}{\rotatebox{90}{Home}} & $H_l @1$ &         0.0049  & 0.0067  & 0.0068  & 0.0066  & 0.0068  & 0.0053  & 0.0049  & 0.0136  & 0.0050  & \underline{0.0142}  & 0.0117  & 0.0128  & \textbf{0.0148*}  \\ 
        ~ & $H_l @3$ & 0.0146  & 0.0150  & 0.0150  & 0.0150  & 0.0169  & 0.0159  & 0.0138  & 0.0247  & 0.0145  & \underline{0.0256}  & 0.0185  & 0.0201  & \textbf{0.0267*}  \\ 
        ~ & $N_l @3$ & 0.0105  & 0.0114  & 0.0114  & 0.0114  & 0.0126  & 0.0114  & 0.0100  & 0.0200  & 0.0104  & \underline{0.0208}  & 0.0156  & 0.0170  & \textbf{0.0217*}  \\ 
        ~ & $H_l @5$ & 0.0222  & 0.0218  & 0.0221  & 0.0217  & 0.0238  & 0.0223  & 0.0217  & 0.0311  & 0.0214  & \underline{0.0325}  & 0.0244  & 0.0256  & \textbf{0.0335*}  \\ 
        ~ & $N_l @5$ & 0.0136  & 0.0142  & 0.0144  & 0.0141  & 0.0155  & 0.0141  & 0.0132  & 0.0226  & 0.0133  & \underline{0.0236}  & 0.0180  & 0.0193  & \textbf{0.0245*}  \\ \midrule

        \multirow{5}{*}{\rotatebox{90}{ML1M}} & $H_l @1$   &      0.0507  & 0.0543  & 0.0514  & 0.0640  & 0.0513  & 0.0640  & 0.0507  & \underline{0.0753}  & - & 0.0717  & 0.0377  & 0.0496  & \textbf{0.0845*}  \\ 
        ~ & $H_l @3$ & 0.1131  & 0.1200  & 0.1120  & 0.1258  & 0.1136  & 0.1337  & 0.1024  & \underline{0.1500}  & - & 0.1487  & 0.0820  & 0.0979  & \textbf{0.1528*}  \\ 
        ~ & $N_l @3$ & 0.0860  & 0.0917  & 0.0860  & 0.0993  & 0.0864  & 0.1039  & 0.0802  & \underline{0.1186}  & - & 0.1154  & 0.0629  & 0.0770  & \textbf{0.1233*}  \\ 
        ~ & $H_l @5$ & 0.1604  & 0.1659  & 0.1586  & 0.1732  & 0.1606  & 0.1838  & 0.1471  & 0.1947  & - & \underline{0.1959}  & 0.1197  & 0.1433  & \textbf{0.2025*}  \\ 
        ~ & $N_l @5$ & 0.1056  & 0.1105  & 0.1051  & 0.1187  & 0.1057  & 0.1245  & 0.0985  & \underline{0.1369}  & - & 0.1347  & 0.0783  & 0.0956  & \textbf{0.1437*}  \\ 
        
        \bottomrule
    \end{tabular}}

\end{table*}

\begin{table}[!ht]
    \setlength{\abovecaptionskip}{0cm}
    \setlength{\belowcaptionskip}{-0.2cm}
    \caption{The Top-1 Generation Accuracy ($H_g@1$) of $SelfDR$ compared with baselines. For traditional recommenders, $H_g@1$ equals $H_l@1$ (both are $HR@1$). *($p$<0.05).}
    \label{tab:gen}
    \centering
    \resizebox{0.85\linewidth}{!}{
    \begin{tabular}{cc|c|c|c}
    \toprule
        \multicolumn{2}{c|}{~} & \textbf{Clothing} & \textbf{Home} & \textbf{ML1M}  \\ \midrule
        \multicolumn{2}{c|}{Base Rank} &  0.0047  & 0.0049  & 0.0507   \\ \midrule
        \multicolumn{2}{c|}{Best ID-Based} & 0.0072  & 0.0068  & \underline{0.0640}  \\  \midrule
        \multicolumn{2}{c|}{Best Content-Based} & 0.0079  & 0.0068  & \underline{0.0640}   \\ \midrule
        \multirow{7}{*}{\parbox{0.8cm}{\centering LLM- \\ Based}} & ZSRanker & 0.0050  & 0.0050  & 0.0466  \\ 
        ~ & SOFT & 0.0097  & 0.0094  & 0.0546  \\ 
        \cmidrule{2-5}
        ~ & EXP3RT & 0.0058  & 0.0050  & - \\ 
        ~ & COT4Rec & 0.0100  & 0.0121  & 0.0575  \\ \cmidrule{2-5}
        ~ & RecSAVER & 0.0108  & 0.0118  & 0.0438  \\ 
        ~ & RecR1 & \underline{0.0114}  & \underline{0.0128}  & 0.0515  \\ \cmidrule{2-5}
        ~ & $SelfDR$ (Ours) & \textbf{0.0118*}  & \textbf{0.0136*}  & \textbf{0.0735*}  \\ 
        \bottomrule
    \end{tabular}}
\end{table}

\subsubsection{Evaluation.} 
We adopt the reranking evaluation setting, where candidates are generated by a base ranker and the positive item is not guaranteed to be included in the candidate set. Compared with ranking over a small set of randomly sampled negatives, this setting better aligns with real-world recommendation scenarios.
To more comprehensively evaluate the LLM performance, we assess its outputs from two perspectives. First, we evaluate the correctness of its generation $s_{u,i}$. Since supervision in LLM training provides only a top-1 label in next-item prediction, generation is typically restricted to a single prediction; thus, we adopt \textit{HitRate@1} as the evaluation metric, denoted as $H_g@1$. Second, we obtain a full ranked list $R_{u,i}$ by sorting candidates according to the log-probabilities of the generated identifier token, and report \textit{HitRate} and \textit{Normalized Discounted Cumulative Gain ($NDCG$)} metric as $H_l$ and $N_l$.

\subsubsection{Baselines.} For a comprehensive evaluation, we compare our method with \textit{\textbf{traditional reranking models}}, including \textit{\textbf{id-based}} (PRM, SetRank, MIR) and \textit{\textbf{content-based}} (BGE, Jina) approaches, as well as \textit{\textbf{LLM-based recommenders}}. Among the LLM-based baselines, \textbf{ZSRanker} and \textbf{SOFT} perform \textit{direct recommendation}, \textbf{EXP3RT} and \textbf{COT4Rec} adopt \textit{multi-step reasoning}, and \textbf{RecSAVER} and \textbf{RecR1} follow the \textit{reasoning-and-recom.} paradigm.

\begin{itemize}
    \item \textbf{PRM}~\cite{PRM}: uses a transformer to model item interactions and user-specific features.
    \item \textbf{SetRank}~\cite{SetRank}: models item interactions for context-aware reranking beyond independent relevance estimation.
    \item \textbf{MIR}~\cite{MIR}: models multi-level interactions across candidate items and user history, employing SLAttention to integrate context for reranking.
    \item \textbf{BGE}~\cite{bge}: a multilingual reranker supporting multi-granular and multi-functional retrieval.
    \item \textbf{Jina}~\cite{jina2024reranker}: a multilingual cross-encoder reranker scoring query–document relevance for retrieval.
    \item \textbf{ZSRanker}~\cite{zsranker}: a prompting-based method using zero-shot LLM with sequential user histories as input.
    \item \textbf{SOFT}~\cite{soft}: a self-optimized fine-tuning method that constructs auxiliary easy-to-learn data with a fine-tuned model to augment real-world data and enhance model training.
    \item \textbf{EXP3RT}~\cite{expert}: a review-based method that trains separate models for user/item profile construction and reasoning generation before prediction. It predicts item ratings and reranks items based on the predicted scores.
    \item \textbf{COT4Rec}~\cite{cot4rec}: a multi-step reasoning method that uses K-means clustering to obtain labels for training an analyzer, then recommends based on the generated analyses.
    \item \textbf{RecSAVER}~\cite{RecSAVER}: a test-time reasoning method that instructs LLMs to generate both reasoning and prediction in one step, with the reasoning label produced by a large LLM and subsequently self-verified. 
    \item \textbf{RecR1}~\cite{recr1}: a test-time reasoning method that directly optimizes LLM generation using feedback from recommendation results through reinforcement learning.
\end{itemize}


\subsubsection{Implementation Details.}

\textbf{Reranking Settings.}
We use SASRec to generate the top-20 candidates in this section. Traditional ID-based baselines are implemented with ReChorus~\cite{rechorus}. All LLM-based recommenders use LLaMA-3.1-8B-Instruct as the backbone, while methods requiring external reasoning labels (\textsc{COT4Rec}, \textsc{EXP3RT}, and \textsc{RecSAVER}) employ the closed-source \textsc{DeepSeek-V3} to generate these labels. Results are averaged over three seeds.

\textbf{Self-Distillation Settings.}
On the training set, we divide the chronologically earlier half for reasoner training and the latter half for distillation. Before distillation, LLMs are instruction-tuned for three epochs to ensure instruction adherence, followed by up to three distillation epochs with early stopping. Since our candidate set size is $20$, we set the $k$ threshold to $5$. Further details on baseline implementation and hyperparameters are provided in our repo.

\subsection{Main Results (RQ1)}


We evaluate LLM-based recommenders from two perspectives. Table~\ref{tab:main} reports the ranking quality of the recommendation lists of our method compared with other baselines, while Table~\ref{tab:gen} focuses on the accuracy of top-1 generation. Since EXP3RT is review-based, it cannot be applied to ML1M, which lacks review data.

\subsubsection{Recommendation List Quality.}
As shown in Table~\ref{tab:main}, our method consistently achieves significantly better performance across all metrics than both traditional and LLM-based recommenders on all three datasets.
This demonstrates the effectiveness and robustness of our approach, highlighting the feasibility of achieving continuous improvement through self-distillation from the reasoning process within a unified model architecture.

For traditional recommenders, most methods yield noticeable gains over the base ranker. ID-based methods, benefiting from task-specific designs, further exploit collaborative-filtering signals and often improve top-ranked accuracy (e.g., @1). Content-based models additionally leverage semantic information from text, which generally strengthens overall ranking quality and brings consistent gains across all cutoffs. Nevertheless, their use of textual signals remains limited, and LLMs, with richer commonsense and world knowledge, have the potential to further improve performance.

For LLM-based methods, appropriate design choices are crucial for eliciting LLMs’ recommendation capability. The performance of ZSRanker suggests that an off-the-shelf 8B model is not sufficiently reliable for direct recommendation. In comparison, SOFT obtains good ranking quality through an enhanced fine-tuning process that leverages both original labels and model-generated curriculum data, although incorporating reasoning still leaves room for further improvement. 
Regarding reasoning-based baselines, EXP3RT improves over the base results only at @1 but degrades at larger cutoffs, as it reranks candidates using LLM-predicted ratings that are often uniformly high, e.g., 4--5, limiting discriminability and list quality. The two reasoning-and-recommendation methods vary markedly across datasets, likely due to differences in textual signals: ML1M mainly provides abstract cues such as movie titles, while Amazon product data contains richer attributes such as gender, capacity and color. This makes reasoning-based learning more challenging on ML1M but relatively easier on Amazon datasets, leading to domain-specific performance variations. COT4Rec is comparatively strong, but its training relies on external-LLM-generated rationales, and such rationales from large closed-source LLMs may not always be recommendation-targeted, introducing noise that leads to suboptimal learning and keeps it below $SelfDR$.

\subsubsection{Top-1 Generation Accuracy.} 
\label{sec:top1}
Beyond recommended lists obtained by log-probability sorting, we also report the LLM's top-1 generation accuracy, denoted as $H_g@1$ (Table~\ref{tab:gen}). For traditional baselines and EXP3RT (which reranks by predicted ratings), $H_g@1$ equals $H_l@1$; therefore, we include only the best method from each traditional category on each dataset for comparison.

Overall, $SelfDR$ consistently achieves the best performance, validating the effectiveness of our method under both ranking-based and generation-based evaluation metrics. Similar to the ranking-quality results, SOFT and COT4Rec still show reasonably good performance, while ZSRanker and EXP3RT remain relatively weak. 

In contrast to the ranking-quality results, the two reasoning-and-recommendation methods achieve strong top-1 generation accuracy on Clothing and Home, especially RecR1. This phenomenon echoes recent observations: reinforcement learning often improves the efficiency of sampling correct outputs, but may not necessarily expand the model’s underlying capability boundary~\cite{rlvr}. Similarly, RecR1 appears more effective at optimizing the generation of the ground-truth token than improving the fine-grained ranking of the ground-truth item among all candidates.

\subsection{Reasoner Performance Comparison (RQ2)}
\label{sec:reasoner}

\subsubsection{Post-Distillation Performance on Test Set}

\begin{figure}[h]
\setlength{\abovecaptionskip}{-0.0cm}
  \centering  \includegraphics[width=\linewidth]{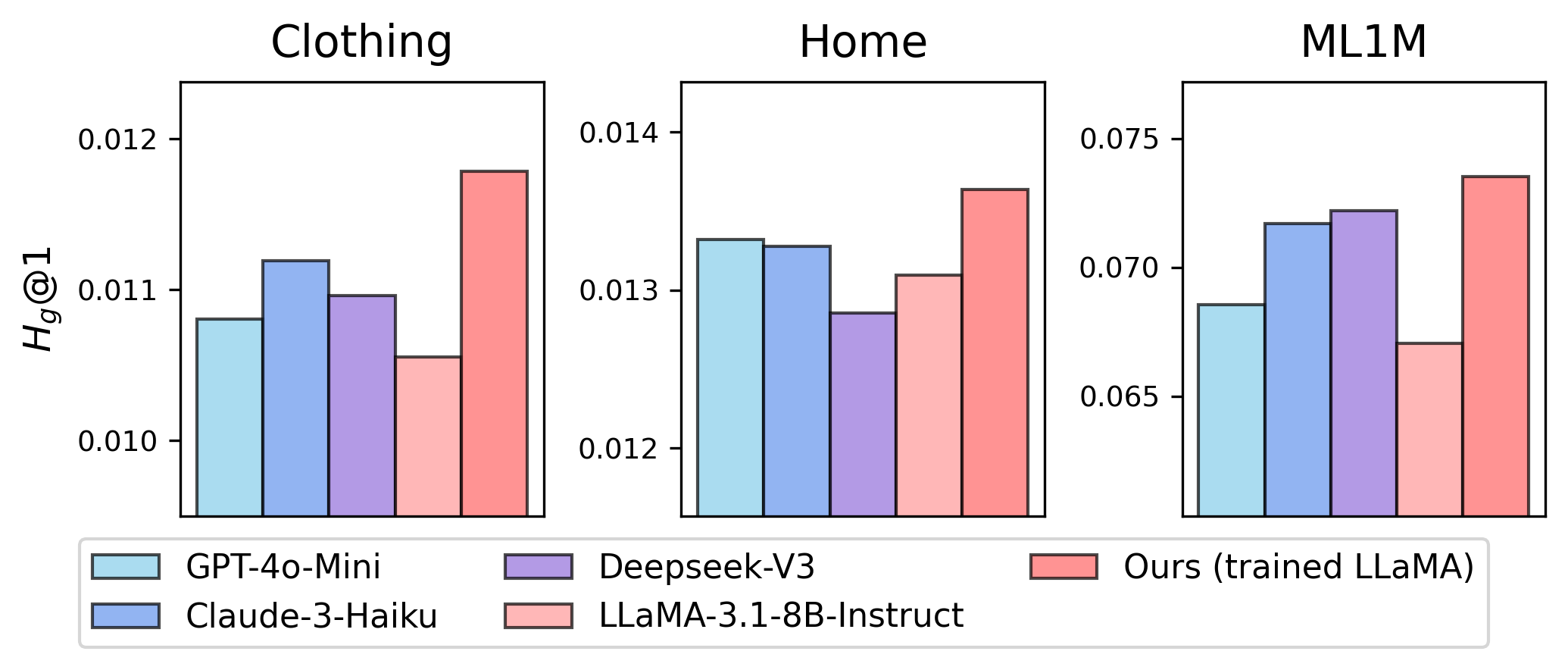}
  \caption{Post-Distillation Recommendation Performance of Different Reasoners (three external large LLMs, untrained LLaMA, and Ours trained LLaMA).}
  \label{fig:reasoner}
\end{figure}

To examine the effectiveness of the $SelfDR$ Reasoner, we conduct experiments where external large LLMs are directly employed as reasoning generators, producing rationales that serve as inputs for the subsequent distillation stage. Specifically, we compare our trained Reasoner with three large LLMs,  \textbf{GPT-4o-mini}, \textbf{Claude-3-Haiku}, and \textbf{DeepSeek-V3}, as well as the \textbf{off-the-shelf LLaMA-3.1-8B-Instruct} baseline.

Figure~\ref{fig:reasoner} reports the final $H_g@1$ results after distillation when using different Reasoners across the three datasets. As shown, our trained Reasoner consistently achieves the best post-distillation performance on all datasets. As expected, the original LLaMA generally leads to worse performance than large proprietary LLMs; however, once trained on the target dataset, it can generate more task-specific and informative explanations that significantly enhance downstream recommendation accuracy after distillation. 
An example of the generated rationales is shown in Sec.~\ref{sec:case}.


Furthermore, we observe that a more powerful general-purpose LLM does not necessarily result in better post-distillation recommendation outcomes. Depending on the dataset, GPT, Claude, or DeepSeek may perform best. This may stem from different domain specialization---each LLM excels in different content domains---or from the cognitive gap between teacher and student models: although large LLMs possess richer knowledge and stronger reasoning abilities, rationales that include knowledge overly unfamiliar or inaccessible to the student may not be easily learnable for a smaller student model.
In contrast, our Reasoner performs consistently well across all datasets, suggesting that using the model itself as the teacher may provide a more balanced and learnable source of supervision for the student.

\subsubsection{Teacher Performance on Training Set}

We further report teacher performance on the training set across different Reasoners, using 2,000 randomly sampled training instances from Clothing due to computational cost.
As shown in Table~\ref{tab:teacher}, all reasoning-enhanced teachers outperform the student before self-distillation regardless of the Reasoner, indicating that rationales generally provide useful auxiliary signals and improve recommendation accuracy. More importantly, this training-set trend largely matches the test-set performance of the distilled students in Figure~\ref{fig:reasoner}: our trained Reasoner performs best, DeepSeek-V3 is competitive among external LLMs, and GPT-4o-Mini performs worst. This consistency suggests that a stronger reasoning-enhanced teacher can offer more effective supervision and better downstream generalization.

In addition, we include GPT-5-Chat as a supplementary observation. Although it outperforms GPT-4o-Mini, it is still not the most effective external Reasoner. This further suggests that a more advanced general-purpose LLM does not necessarily yield better downstream performance, possibly due to the mismatch between external large LLMs and the smaller recommender LLM in reasoning style, output distribution, or recommendation-specific alignment.

\begin{table}[!ht]
\setlength{\abovecaptionskip}{0cm}
\setlength{\belowcaptionskip}{0cm}
\caption{Teacher Performance on the Clothing training set when constructed with rationales generated by different Reasoners on 2,000 randomly sampled training instances.}
\label{tab:teacher}
    \centering
    \resizebox{\linewidth}{!}{
    \begin{tabular}{c|c|ccccc}
    \toprule
        Reasoner & $H_g@1$ & $H_l@1 $& $H_l@3$ & $N_l@3$ & $H_l@5$ & $N_l@5$  \\ \midrule 
        \parbox{3cm}{\centering No (Student Before SD)} & 0.373  & 0.454  & 0.763  & 0.6336  & 0.876  & 0.6802  \\ \midrule
        LLaMA & 0.407  & 0.503  & 0.825  & 0.6916  & 0.911  & 0.7272  \\ \midrule
        GPT-4o-Mini & 0.416  & 0.505  & 0.820  & 0.6896  & 0.911  & 0.7271  \\ 
        Claude-3-Haiku & \underline{0.508}  & \underline{0.601}  & \underline{0.901}  & \underline{0.7805}  & \underline{0.963}  & \underline{0.8061}  \\ 
        Deepseek-V3 & 0.478  & 0.572  & 0.848  & 0.7345  & 0.930  & 0.7684  \\ 
        GPT-5-Chat & 0.429  & 0.508  & 0.809  & 0.6838  & 0.909  & 0.7253  \\ \midrule
        \textit{Ours (trained LLaMA)} & \textbf{0.552}  & \textbf{0.615}  & \textbf{0.910}  & \textbf{0.7908}  & \textbf{0.968}  & \textbf{0.8151}  \\ 
        \bottomrule
    \end{tabular}
    }
\end{table}

\subsubsection{Qualitative Comparison of Reasoner-Generated Rationales}
\label{sec:case}
An example of the rationales generated by different Reasoners on Clothing is shown in Table~\ref{tab:case}. By summarizing the user’s historical interactions, we observe several key preferences: the items are generally functional and stylish, and some of them are purchased as gifts. However, given the next item, the more relevant signals in this case are functionality and gift-oriented intent. The off-the-shelf LLaMA captures most characteristics from the user history, but fails to emphasize the aspects most relevant to the target item. Claude better connects the user history with the next item, yet still misses the important gift-related signal. In contrast, our trained Reasoner generates a rationale with the richest and most accurate information. This demonstrates its stronger ability to produce high-quality rationales, which further contributes to more effective teacher construction and better post-distillation performance.

\definecolor{deepblue}{RGB}{0,60,130}
\definecolor{darkred}{RGB}{150,20,20}
\definecolor{darkochre}{RGB}{160,105,25}
\begin{table}[!ht]
\setlength{\abovecaptionskip}{0cm}
\setlength{\belowcaptionskip}{-0.2cm}
\caption{Rationales of one training instance on Clothing generated by various Reasoners. Same-colored texts convey similar points, with \textit{Ours} being more targeted and informative.}
\label{tab:case}
    \centering
    \resizebox{\linewidth}{!}{
    \begin{tabular}{c|p{8.5cm}} 
    \toprule
        \textbf{Reasoner} & \textbf{Generated Rationale}  \\ \midrule 
        LLaMA & Based on the user's past purchase history, it can be inferred that they tend to purchase items related to \textbf{\textcolor{darkochre}{fashion ... values high-quality, stylish items}} ... open to \textbf{\textcolor{darkred}{purchasing items as gifts for others}}. ... suggest that the user values \textbf{\textcolor{deepblue}{practical}} or \textbf{\textcolor{darkochre}{stylish}} items that would be appreciated by others. \par The new item ... is a \textbf{\textcolor{deepblue}{casual and versatile}} piece of clothing, may also appeal to the user's taste in \textbf{\textcolor{deepblue}{practical}} and \textbf{\textcolor{darkochre}{stylish}} items. \\ \midrule
        Claude & They seem to have a preference for \textbf{\textcolor{deepblue}{versatile, everyday clothing items that are practical and functional}}. The high rating they gave to ... suggests they value products that can be easily incorporated into their daily wardrobe. \par The new item aligns well with the user's apparent preference for \textbf{\textcolor{deepblue}{simple, comfortable, and easy-to-wear clothing}}...  \\ \midrule
        \textit{Ours} & We can infer that the user tends to favor products that are \textbf{\textcolor{deepblue}{functional}} and \textbf{\textcolor{darkred}{perhaps given as gifts}}. They purchased ... suggests that they appreciate quality and functionality. \par The new item shares similarities with the user's past purchase. It is a \textbf{\textcolor{deepblue}{practical, everyday item of clothing that can be worn in casual settings}}... Additionally, ... suggests that the user may be \textbf{\textcolor{darkred}{purchasing this item for a younger family member or friend}}, aligning with their past behavior of giving gifts.  \\ \bottomrule
    \end{tabular}}
\end{table}

\subsection{Effect of Self-Distillation (RQ3)}

\subsubsection{Ablation on Self-Distillation}
In this section, we conduct ablation experiments to validate the contribution of each component in the self-distillation stage. Specifically, we compare four settings: (1) \textbf{Before Self-Distillation \textit{(Before SD)}}, where the model is only instruction-tuned without any further self-distillation;
(2) \textbf{Supervised-only training \textit{(Sup-Only)}}, where the self-distillation stage uses only the cross-entropy loss;
(3) \textbf{Distillation-only training \textit{(Dist-Only)}}, which applies only the knowledge distillation loss; and
(4) \textbf{Fixed-$\alpha$}, where dynamic weighting is disabled and $\alpha$ is fixed to 0.5.
For all settings involving self-distillation, we keep the number of training epochs consistent to ensure a fair comparison.

\begin{table}[!t]
\setlength{\abovecaptionskip}{0cm}
\setlength{\belowcaptionskip}{0cm}
\caption{The Results of $SelfDR$ with Dynamic Weighting compared with other learning strategies on Clothing. Bold and underlined indicate the best and the second-best.}
\label{tab:curriculum}
    \centering
    \resizebox{\linewidth}{!}{
    \begin{tabular}{c|c|ccccc}
    \toprule
        ~ & $H_g@1$ & $H_l@1 $& $H_l@3$ & $N_l@3$ & $H_l@5$ & $N_l@5$  \\ \midrule
        Before SD & 0.0094  & 0.0112  & 0.0240  & 0.0186  & 0.0311  & 0.0215  \\ \midrule
        Sup-Only & 0.0092  & 0.0108  & 0.0236  & 0.0182  & 0.0313  & 0.0214  \\
        Dist-Only & \underline{0.0115}  & \textbf{0.0132}  & \underline{0.0247}  & \underline{0.0198}  & 0.0314  & \underline{0.0226}  \\ 
        Fixed-$\alpha$ & 0.0109  & 0.0124  & 0.0245  & 0.0194  & \underline{0.0315}  & 0.0223  \\ \midrule
        $SelfDR$ & \textbf{0.0118}  & \textbf{0.0132}  & \textbf{0.0249}  & \textbf{0.0200}  & \textbf{0.0318}  & \textbf{0.0228}  \\ \bottomrule
    \end{tabular}
    }
\end{table}

Table~\ref{tab:curriculum} reports the recommendation results on Clothing, where our method achieves significantly better overall performance.
We observe that continuing training with only the supervised CE loss tends to overfit the training data and degrades generalization. In contrast, distilling from the reasoning-augmented teacher provides additional informative signals for the student and yields further gains. However, because the teacher’s predictions are not always perfectly accurate, introducing supervision from the ground-truth labels helps stabilize learning and improves robustness. Notably, this combination requires careful design: a naïve mixture of distillation and label supervision can underperform distillation-only training, whereas our dynamic weighting strategy enables a more effective balance between the two.

\subsubsection{Impact of Self-Distillation on Output Confidence}

To better understand how self-distillation affects the LLM, we further examine the token-level probability distribution of its outputs. We group model outputs according to the position of the ground-truth identifier in the generated recommendation list $\mathcal{R}_{u,i}$, and visualize the probability distribution, i.e., the model's confidence, assigned to the ground-truth identifier within each group. 

\begin{figure}[h]
\setlength{\abovecaptionskip}{-0.0cm}
\setlength{\belowcaptionskip}{-0.0cm}
  \centering  \includegraphics[width=\linewidth]{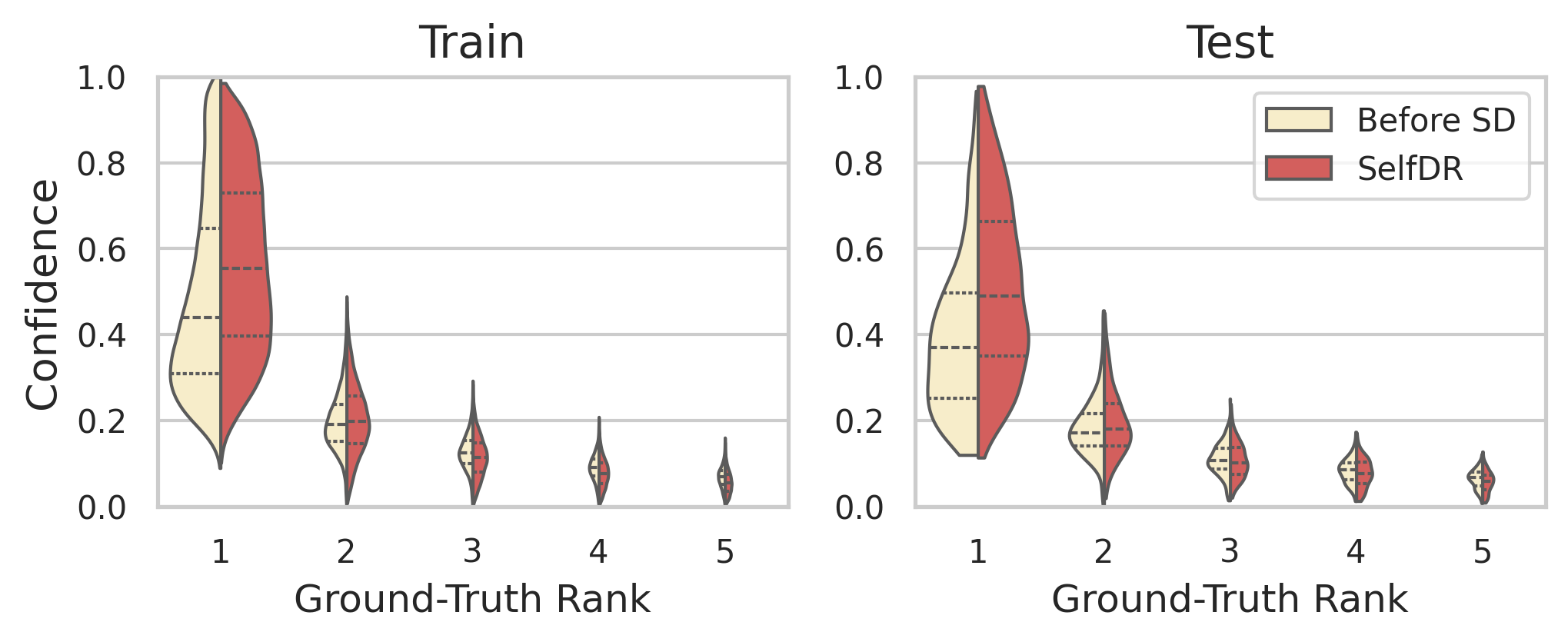}
  \caption{Confidence of the Ground-Truth Identifier ranked within the top-5 on Clothing. The left and right panels show the distributions on the training and test sets, respectively.}
  \label{fig:analysis}
  \vspace{-10pt}
\end{figure}

Figure~\ref{fig:analysis} reports the confidence distributions when the ground-truth identifier appears within the top-5 positions on the training and test sets of Clothing. In each violin plot, wider regions indicate that more samples are concentrated around the corresponding confidence values.
As shown, SelfDR substantially increases the model’s confidence when the ground-truth item is ranked within the top-2 positions. This indicates that, after self-distillation, the model not only ranks correct items higher, as reflected in improved ranking metrics such as \textit{NDCG}, but also becomes more confident in correct cases, both of which contribute to more accurate generation during inference. Moreover, this pattern remains consistent across the training and test sets, suggesting that the confidence adjustment is stable rather than an artifact of the training distribution, and providing evidence that the student model internalizes transferable ranking knowledge through self-distillation.

\subsection{Training Cost \& Inference Efficiency (RQ4)}

To estimate the training cost and inference efficiency of LLM-based recommenders, we run both training and inference on a single NVIDIA A800 GPU. Training time is reported as the total wall-clock time required to obtain all models used in the final setup. For inference, we randomly sample 2,000 instances from Clothing and report the per-instance average inference time and token usage. 

\begin{table}[!ht]
    \setlength{\abovecaptionskip}{0cm}
    \setlength{\belowcaptionskip}{-0.2cm}
    \caption{Training Cost and Inference Efficiency on Clothing (training time in hours; inference time in ms/instance). For EXP3RT and COT4Rec, totals across all steps are reported.}
    \label{tab:inference}
    \centering
    \resizebox{\linewidth}{!}{
    \begin{tabular}{c|cccc}
    \toprule
        ~ & \textbf{Input Tok.} & \textbf{Output Tok.} & \textbf{Train. Time} & \textbf{Inf. Time}\\ \midrule
        ZSRanker & 1253.13  & \underline{24.07}  & - & 149.81  \\ 
        SOFT & \underline{1198.06}  & \textbf{1.00}  & \textbf{37.79} & \underline{109.48}  \\ \midrule
        EXP3RT & 1972.40  & 670.01 & 47.65 & 529.83  \\ 
        COT4Rec & 1511.87 & 168.36 & 58.36 & 250.03  \\ \midrule
        RecSAVER & 1472.21  & 197.51 & \underline{40.11} & 623.38  \\ 
        RecR1 & \textbf{1182.06 } & 210.43 & 65.50 & 665.12  \\ \midrule
        \textbf{$SelfDR$} & \underline{1198.06}  & \textbf{1.00} & 54.18 & \textbf{109.46}  \\ 
        \bottomrule
    \end{tabular}}
\end{table}


As shown in Table~\ref{tab:inference}, $SelfDR$ achieves the most efficient inference, tied with SOFT, by producing only a single-character identifier, while keeping its training time comparable to most reasoning-based baselines without incurring a dramatic increase in training overhead. On the inference side, SOFT, which does not involve reasoning, achieves efficiency close to ours; however, all other baselines are substantially slower. Although ZSRanker does not generate rationales, it still needs to output both identifiers and titles to prevent the zero-shot LLM from drifting to arbitrary answers, thereby increasing latency. Reasoning-based methods, whether multi-step or test-time approaches, inevitably introduce additional inference overhead. EXP3RT partially reduces per-step latency through shorter inputs and outputs, but its multi-stage pipeline still leads to higher overall inference cost.
In terms of training, our reasoner converges quickly, and the recommender’s one-token output keeps both instruction tuning and self-distillation lightweight. Meanwhile, existing baselines often train multiple models or generate longer sequences, both of which increase training overhead. Consequently, despite using a two-stage pipeline, $SelfDR$ maintains a total training time comparable to prior methods and is notably more efficient than RecR1, which requires a substantially longer GRPO procedure. Moreover, $SelfDR$ does not rely on external LLMs during training, whereas EXP3RT, COT4Rec, and RecSAVER incur additional costs by querying state-of-the-art LLMs for supervision signals.


\section{Discussions}

\subsection{Hyperparameter Sensitivity Analysis}
\label{sec:hyper}

\begin{figure}[h]
\setlength{\abovecaptionskip}{-0.0cm}
  \centering  \includegraphics[width=0.97\linewidth]{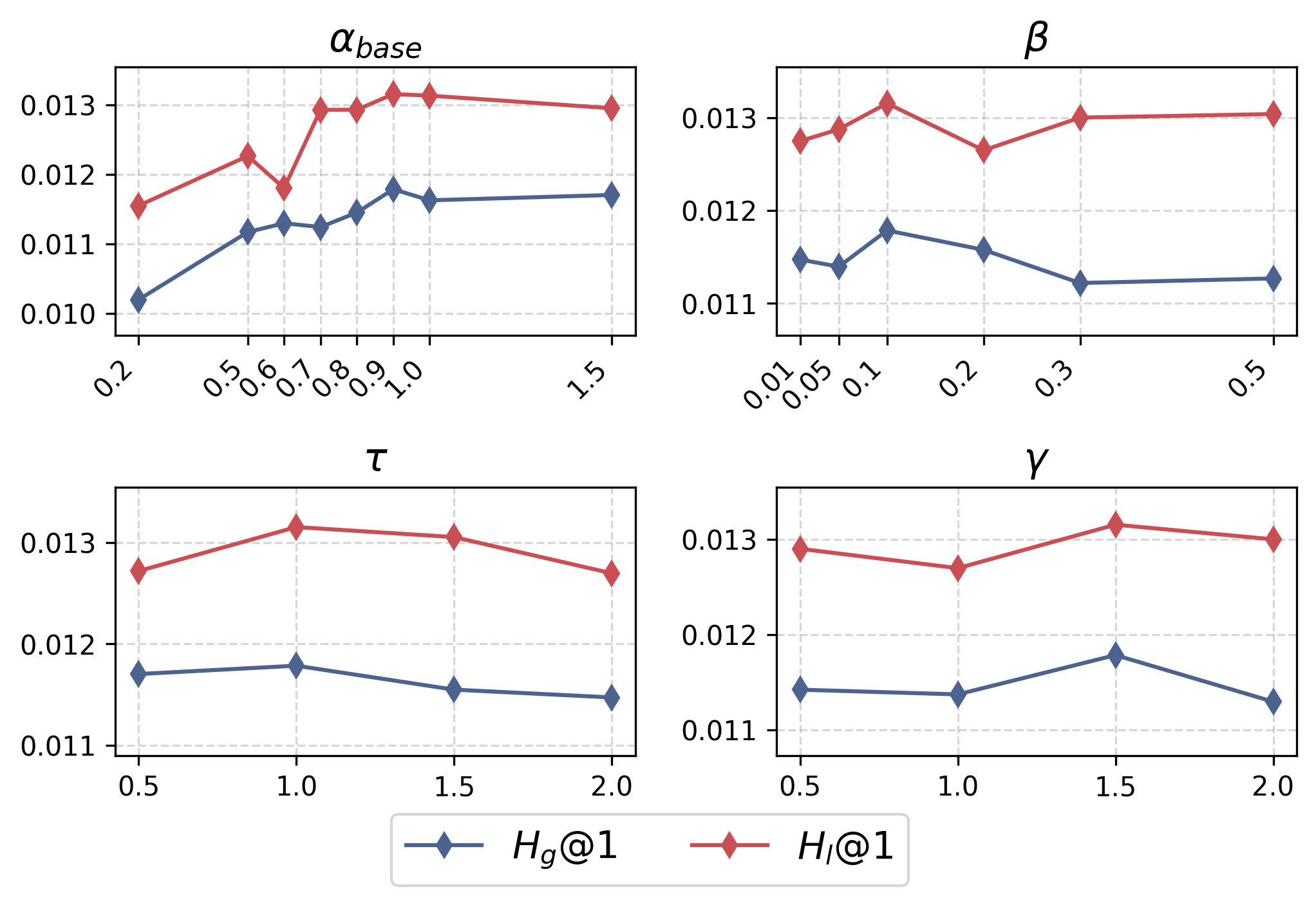}
  \caption{Hyperparameter Analysis of the dynamic weighting strategy on Clothing. For each hyperparameter, we report its impact on $H_g@1$ and $H_l@1$.}
  \label{fig:hyper}
  \vspace{-10pt}
\end{figure}

In this section, we study the sensitivity of the dynamic weighting strategy to $\alpha_{\text{base}}$, $\beta$, and the temperature coefficients $\tau$ and $\gamma$. Figure~\ref{fig:hyper} reports results on Clothing, with similar trends on the other two datasets. Overall, the optimal $\alpha_{\text{base}}$ varies across datasets, $\beta\approx0.1$ usually performs best, and the temperature coefficients have relatively minor effects, with best results typically in $[1.0, 2.0]$.

Specifically, on Clothing, $\alpha_{\text{base}}=0.9$ achieves the best overall performance. A smaller $\alpha_{\text{base}}$ pushes training toward supervised fine-tuning, increasing overfitting and causing a clear performance drop, while an overly large $\alpha_{\text{base}}$ pushes it toward a distillation-dominated regime near $\alpha_{\max}$. Although stronger self-distillation can still be competitive when $\beta$ properly suppresses noisy supervision, a moderate amount of cross-entropy supervision remains beneficial.
The parameter $\beta$ down-weights instances where the teacher underperforms the student, and $\beta=0.1$ usually provides effective separation. A larger $\beta$ weakens this penalty and admits more noisy teacher signals, while a very small $\beta$ makes such instances rely mostly on label supervision. Even so, a worse-ranked teacher's soft token-level distribution may still offer useful information beyond a hard label.
The model is generally robust to the temperature coefficients. Since these two parameters control how the teacher rank and rank difference affect the adaptive weight $\alpha$, a moderate range suffices for strong performance. Overall, these results show that $SelfDR$ is stable under reasonable hyperparameter choices, confirming the robustness of the dynamic weighting strategy.

\subsection{Candidate Generator Robustness}

\begin{table}[!ht]
    \setlength{\abovecaptionskip}{0cm}
    \setlength{\belowcaptionskip}{0cm}
    \caption{Recommendation Performance of $SelfDR$ compared with the strongest baselines on Clothing using LightGCN and ComiRec as the candidate generator. *($p$<0.05).}
    \label{tab:generator}
    \centering
    \resizebox{\linewidth}{!}{
    \begin{tabular}{c|c|ccccc}
    \toprule
    LightGCN & $H_g@1$ & $H_l@1 $& $H_l@3$ & $N_l@3$ & $H_l@5$ & $N_l@5$ \\ \midrule
        Base Rank & 0.0066  & 0.0066  & 0.0161  & 0.0120  & 0.0235  & 0.0150  \\ \midrule
        BGE & \underline{0.0094}  & 0.0094  & 0.0212  & 0.0162  & 0.0292  & 0.0195  \\ 
        COT4Rec & 0.0089  & \underline{0.0107}  & \underline{0.0232}  & \underline{0.0178}  & \underline{0.0318}  & \underline{0.0214}  \\ \midrule
        $SelfDR$ & \textbf{0.0099*}  & \textbf{0.0110*}  & \textbf{0.0245*}  & \textbf{0.0188*}  & \textbf{0.0328*}  & \textbf{0.0221*}  \\ \midrule
    ComiRec & $H_g@1$ & $H_l@1 $& $H_l@3$ & $N_l@3$ & $H_l@5$ & $N_l@5$ \\ \midrule
        Base Rank & 0.0068  & 0.0068  & 0.0183  & 0.0134  & 0.0264  & 0.0168  \\ \midrule
        BGE & 0.0121  & 0.0121  & 0.0257  & 0.0200  & 0.0344  & 0.0235  \\ 
        COT4Rec & \underline{0.0131}  & \underline{0.0158}  & \underline{0.0316}  & \underline{0.0248}  & \underline{0.0398}  & \underline{0.0282}  \\ \midrule
        $SelfDR$ & \textbf{0.0167*}  & \textbf{0.0192*}  & \textbf{0.0346*}  & \textbf{0.0281*}  & \textbf{0.0429*}  & \textbf{0.0315*}  \\ 

    \bottomrule
    \end{tabular}}
\end{table}

In the experiments above, we have shown that our method consistently achieves superior performance over various baselines across multiple datasets, with SASRec serving as the base ranker. We further investigate whether this advantage remains robust when the candidate generator changes. To this end, we conduct additional experiments on Clothing using two different base ranker: (1) LightGCN~\cite{LightGCN}, a graph-based collaborative filtering model that captures high-order user--item relations; (2) ComiRec~\cite{ComiRec}, a sequential recommendation model that captures multiple user interests from behavior sequences. We compare $SelfDR$ with BGE and COT4Rec, which are the strongest baselines overall among traditional and LLM-based methods, respectively.
As shown in Table~\ref{tab:generator}, all three methods improve over the original base ranker. Moreover, $SelfDR$ consistently achieves the best performance across all metrics, demonstrating its robustness and generalizability across different candidate generators, as well as its stable effectiveness on the task.

\subsection{Out-of-Domain Generalization}

To verify that self-distillation learns transferable knowledge rather than dataset-specific shortcuts, we conduct an out-of-domain evaluation. Specifically, we take models trained on one dataset, before and after self-distillation, and directly evaluate them on other datasets. Improvements on unseen domains would indicate that the gains are not merely due to memorizing superficial patterns in the training data.
Table~\ref{tab:ood} reports the results of models trained on Clothing and Home and evaluated on the other dataset. The distilled model consistently outperforms its non-distilled counterpart in both transfer directions, demonstrating that the benefits of self-distillation generalize beyond the training domain. These consistent gains suggest that self-distillation helps the model internalize more transferable recommendation knowledge, rather than simply fitting dataset-specific signals.


\begin{table}[!ht]
    \setlength{\abovecaptionskip}{0cm}
    \setlength{\belowcaptionskip}{-0.2cm}
    \caption{Out-of-Domain Performance before and after self-distillation. The upper block evaluates Clothing-trained models on Home, while the lower block evaluates Home-trained models on Clothing.}
    \label{tab:ood}
    \centering
    \resizebox{\linewidth}{!}{
    \begin{tabular}{l|c|ccccc}
        \toprule
        Dat: Home & $H_g@1$ & $H_l@1 $& $H_l@3$ & $N_l@3$ & $H_l@5$ & $N_l@5$ \\ \midrule
        Before SD & 0.0076  & 0.0090  & 0.0197  & 0.0152  & 0.0275  & 0.0184  \\ 
        $SelfDR$ & \textbf{0.0095}  & \textbf{0.0106}  & \textbf{0.0208}  & \textbf{0.0165}  & \textbf{0.0279}  & \textbf{0.0194}  \\ \midrule
        Dat: Clothing & $H_g@1$ & $H_l@1 $& $H_l@3$ & $N_l@3$ & $H_l@5$ & $N_l@5$ \\ \midrule
        Before SD & 0.0100  & \textbf{0.0127}  & 0.0237  & 0.0191  & 0.0312  & 0.0222  \\
        $SelfDR$ & \textbf{0.0115}  & \textbf{0.0127}  & \textbf{0.0243}  & \textbf{0.0194}  & \textbf{0.0322}  & \textbf{0.0227}  \\  
        \bottomrule
    \end{tabular}}
\end{table}


\section{Conclusion and Future Work}

In this paper, we introduce $SelfDR$, a novel self-distillation from reasoning framework for LLM-based recommendation.
$SelfDR$ self-distills the LLM’s own reasoning-enhanced predictions to achieve self-learning and evolution, enhancing recommendation effectiveness while maintaining efficiency.
Specifically, $SelfDR$ first builds a reasoning-guided teacher by training a reasoner to generate targeted rationales, and then lets a student model with the same architecture learn from the teacher’s deliberated outputs through dynamic loss.
Experiments on three public datasets show that $SelfDR$ achieves the best recommendation effectiveness while maintaining high inference efficiency. 
Ablation studies further verify that a compact LLM itself, once trained as the reasoner, can outperform external large LLMs, and that the dynamic self-distillation design is key to stable and effective learning.
This work represents an early exploration of distilling from the reasoning of LLMs to realize self-evolution within a unified architecture.
Our work centers on the widely-used LLaMA-8B model, aiming to enable relatively modest models to self-enhance their recommendation capabilities.
Future work may extend this paradigm to LLMs of different types and scales to further validate its adaptability, moving toward autonomously evolving LLMs that continuously refine their reasoning and decision-making capabilities without external teachers.


\clearpage
\section*{GenAI Usage Disclosure} 

The authors used generative AI tools only for language editing and polishing of author-written text, such as improving grammar, clarity, and fluency. These tools were not used to generate research ideas, methods, data, experimental results, citations, or scientific claims. All AI-assisted edits were carefully reviewed, verified, and revised by the authors. The authors take full responsibility for the accuracy, originality, and integrity of all content in this work.
\bibliographystyle{ACM-Reference-Format}
\bibliography{reference}


\end{document}